# A Route To High Performance Micro-Solid Oxide Fuel Cells On Metallic Substrates

*Matthew P. Wells[1], Adam J. Lovett[1], Thomas Chalken[1], Federico Baiutti[2], Albert Tarancón,[2,3] Xuejing Wang[4], Jie Ding[4], Haiyan Wang[4], Sohini Kar-Narayan[1], Matias Acosta[1], Judith L. MacManus-Driscoll[1]*

[1]Department of Materials Science and Metallurgy, University of Cambridge, 27 Charles Babbage Road, Cambridge CB3 0FS, United Kingdom.

[2]Catalonia Institute for Energy Research (IREC), Department of Advanced Materials for Energy, 1 Jardins de les Dones de Negre, Barcelona 08930, Spain.

[3]ICREA, 23 Passeig Lluís Companys, Barcelona 08010, Spain

[4]School of Materials Engineering, Purdue University, 701 West Stadium Avenue, West Lafayette 47907-2045, United States.

KEYWORDS. Solid oxide fuel cell; epitaxial thin film; metallic substrate; commercially viable; high-performance.

ABSTRACT

Micro-solid oxide fuel cells based on thin films have strong potential for use in portable power devices. However, devices based on silicon substrate typically involves thin-film metallic electrodes which are unstable at high temperatures. Devices based on bulk metal substrates overcome these limitations, though performance is hindered by the challenge of growing state-of-the-art epitaxial materials on metals. Here, we demonstrate for the first time the growth of epitaxial cathode materials on metal substrates (stainless steel), commercially supplied with epitaxial electrolyte layers (1.5 μm $(Y_2O_3)_{0.15}(ZrO_2)_{0.85}$ (YSZ) + 50 nm $CeO_2$). We create epitaxial mesoporous cathodes of $(La_{0.60}Sr_{0.40})_{0.95}Co_{0.20}Fe_{0.80}O_3$ (LSCF) on the substrate by growing LSCF /MgO vertically aligned nanocomposite (VAN) films by pulsed laser deposition (PLD), followed by selectively etching out the MgO. To enable valid comparison with literature, the cathodes are also grown on single-crystal substrates, confirming state-of-the-art performance with an area specific resistance of 100 Ω at 500° C and activation energy down to 0.97 eV. The work marks an important step towards the commercialisation of high-performance micro-solid oxide fuel cells for portable power applications.

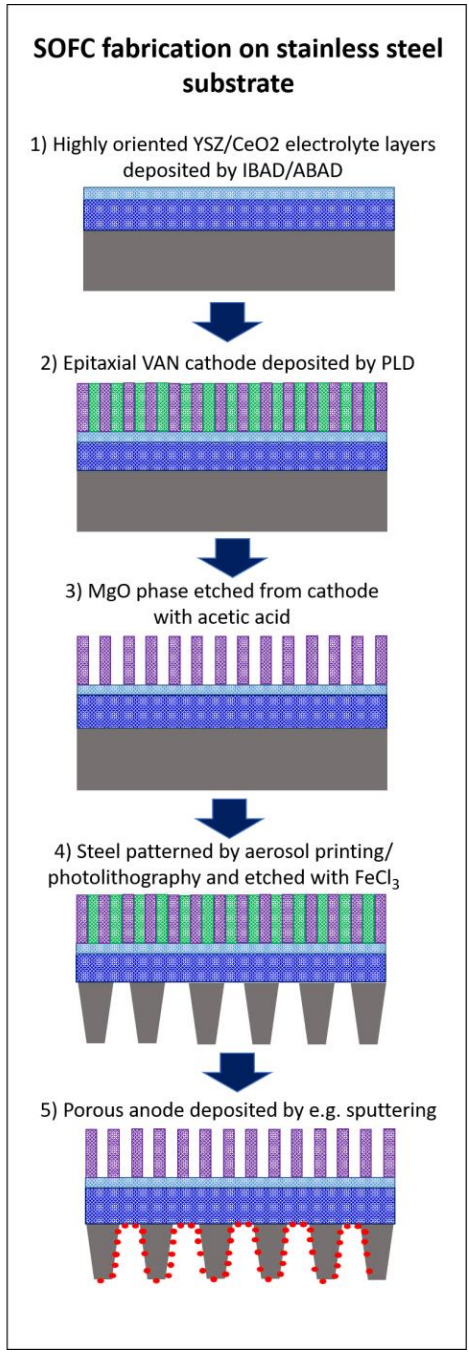

Fig. 1: Schematic of fabrication procedure for micro-solid oxide fuel cells incorporating vertically aligned nanocomposite films on stainless steel substrates

1. INTRODUCTION

Micro-solid oxide fuel cells (μSOFCs) are considered a promising future technology for portable power applications due to their efficiency and high energy densities (up to 4 times higher than Li-ion batteries) [1]. To achieve maximum efficiency, bulk SOFCs operate at temperatures in the region of 800° C [2]. This, however, leads to both long start-up times and a lack of suitability to small-scale portable power applications, where devices require operation temperatures well below 500° C.

Despite the many advances of μSOFCs [1-3], two key challenges remain a) reducing the cell area specific resistances (ASRs), particularly the cathode ASR, so that cells can be operated below 500° C, and b) improving device stability with temperature cycling. While μSOFCs on Si [4, 5], a commercially viable substrate with ease of device integration [6, 7], show good promise, there are problems of cyclability. Even when operating in the low-temperature regime of SOFC's (300 to 500° C), Si-based devices are susceptible to strong degradation mainly due to the instability of thin-film metallic electrodes at intermediate to high temperature ranges [2, 8-10].

A possible strategy for achieving low ASRs at <500° C, while maintaining a good in-plane electrical conductivity for current collection and compatibility with thin film technology, is the use of textured or epitaxial thin film cathodes and electrolytes of low thickness[11]. Particularly, textured or epitaxial layers offer the advantage of a well-controlled geometry and microstructure, whether for the electrolyte, cathode or both [12-14]. The highest power loss in a μSOFC is typically due to the oxygen reduction reaction (ORR) overpotential at the cathode and there are therefore many reports of state-of-the-art epitaxial cathode thin films [6, 15], [12, 16]. Primarily μSOFCs have been fabricated on Si, but research has also focused on μSOFC fabrication on metal substrates to enable fast start-up times and superior mechanical stability and cyclability [2, 17, 18], as discussed later.

As far as lowering ASR of the electrolyte goes control of grain boundaries and porosity in textured YSZ films has yielded values for activation energy and ionic conductivity of 1.04 eV and 0.02 S/m at 500 °C respectively. These values are lower those of the bulk material for which respective values of 1.18 eV and 0.1 S/m are reported [19]. Further, epitaxial YSZ films on Si have shown activation energy of 0.79 eV and ionic conductivity of ~0.003 S/m at 500 °C marking further performance enhancements compared to textured films[20]. Exceptional improvements on electrolyte performance have also been achieved using vertically aligned nanocomposite (VAN) systems. More than an order of magnitude ionic conductivity was shown in several thin film VAN systems (YSZ, $SrZO_3$, and Sm-doped $CeO_2$) [21-23], pointing towards the recent concept of room temperature electrolytes [24]. However, to date, the superior performance of VAN films has only been demonstrated on single crystal substrates.

As far as reducing cathode ASR values and activation energies go, epitaxial films of state-of-the-art cathode materials such as $(La_{0.60}Sr_{0.40})_{0.95}Co_{0.20}Fe_{0.80}O_3$ (LSCF) and $La_{0.7}Sr_{0.3}MnO_3$ (LSM), are again typically grown on well lattice-matched single crystal STO, $NdGaO_3$ and $LaAlO_3$ substrates [12, 16]. At 500° C, ASR values for LSCF are reported in the region of 10 $\Omega cm^2$ with an activation energy between 1.2 and 1.7 eV [25]. Such epitaxial cathodes allow precise control of crystallographic planes at the surface, allowing substantial modification to reaction kinetics[26, 27]. Meanwhile, VAN cathodes enable maximisation of the reaction area and triple phase boundary[28]. [29]

Fabricating μSOFCs on low-cost and flexible metal substrates such as Ni or stainless steel, with thermal expansion coefficients closely matching YSZ, overcomes the issues related to the use of instable thin film metals in Si as well as the very different thermal expansions between the substrate and the electrolyte layer. Metal substrates also provide mechanical strength, high electrical conductivity, and good thermal conduction, enabling quick start-up times [2, 17, 18]. However, metal substrates tend to preclude the use of advanced, epitaxially grown, electrolyte and cathode materials. To our knowledge, no reports describe the growth of epitaxial electrolytes and cathodes on metal substrates. This is an essential challenge which must be overcome to enable the proliferation of high-performance, low temperature μSOFCs, and this forms the focus of this study.

In the present work, we explore the use of metal substrates on which are grown highly biaxially oriented thin films of YSZ (1500 nm) and $CeO_2$ (50nm) made using ion beam or alternating beam assisted deposition (IBAD/ABAD), a scalable and cost-effective technique which have already found commercial applicability for the deposition of YSZ films in the superconductor industry [30, 31]. We follow a μSOFC device fabrication procedure as set out in Fig. 1. We show that mesoporous VAN LSCF cathodes of high quality can be easily grown on stainless steel substrates with biaxially textured YSZ electrolyte-$CeO_2$ bilayers prepared by ABAD. It may be noted that, while strained $CeO_2$ shows good ionic conductivity, further performance enhancements can readily be achieved by doping [32, 33]. We also demonstrate a simple means to chemically etching the stainless steel to allow gas access at the anode. Impedance measurements show that this μSOFC system is well-suited to low temperature operation due to a low activation energy resulting in low cathode area specific resistance (ASR) in the 300° C to 500° C temperature range. The present work marks an important step towards the commercialisation of high-performance μSOFCs for portable power applications.

2. EXPERIMENTAL SECTION

Before determining the success of the μSOFC structure and properties, it was necessary to first assess the properties of the 'ideal' VAN cathode of $(La_{0.60}Sr_{0.40})_{0.95}Co_{0.20}Fe_{0.80}O_3$ (LSCF)/MgO grown on single crystal $(Y_2O_3)_{0.08}(ZrO_2)_{0.92}$ (YSZ). We also grew films on single crystal LSAT to determine whether a mesoporous honeycomb structure is not highly susceptible to lattice or structural mismatch effects and is hence a more universal approach which could later be applied to growth on a range of different electrolyte systems.

Cathode films were grown by pulsed laser deposition (PLD) on (001) oriented $(Y_2O_3)_{0.08}(ZrO_2)_{0.92}$ (YSZ) and $(LaAlO_3)_{0.3}(Sr_2TaAlO_6)_{0.7}$ (LSAT) substrates (and later on metal substrates as outlined below). The LSCF/MgO targets for the PLD growth were made by mixing 50:50 wt.% $(La_{0.60}Sr_{0.40})_{0.95}Co_{0.20}Fe_{0.80}O_3$ (Fuel cell materials, Nexceris, LLC) and MgO powders (Alfa Aesar) in a rotary ball mill for ~15hrs. The mixed powders were pelletised into discs 12 mm in diameter and approximately 2 mm thick. The pellets were then placed in a 1500 bar cold isostatic press before sintering at 1300° C for 4 hours (heated and cooled at 5° C/min).

Films were deposited at 750° C in 0.4 mBar $O_2$ atmosphere with a flow rate of 9.8 sccm. The chamber was evacuated to at least $10^{-7}$ mBar before deposition. A composite ceramic target, held 45 mm from the substrates, was ablated with a 248 nm laser (Lambda Physik, Inc) with a repetition rate of 3 Hz and a fluence of approximately 0.8 J/cm$^2$. Samples were cooled to room temperature in the same 0.4 mBar $O_2$ atmosphere at 10° C/min.

The procedure for the fabrication of the μSOFCs (shown schematically in Fig. 1) is as follows:

**Step 1 - metal/electrolyte substrates:** Commercial (Bruker) stainless steel substrates (0.1mm) were sourced with biaxially textured YSZ (1500 nm) and $CeO_2$ (50 nm) layers prepared by ABAD.

**Step 2 - deposition of VAN cathodes**: VAN cathode films were deposited on the ABAD-buffered stainless steel substrates by PLD using the same parameters described above for the single crystal substrates above.

**Step 3 - etching the VAN cathodes:** MgO was etched out of the VAN film using a 3.3 M solution of acetic acid, stirred at 60 rpm for 17 minutes, to leave a mesoporous LSCF cathode.

**Step 4 - etching holes in stainless steel:** The stainless steel was patterned to give a mask of ~100 μm diameter holes spaced ~500 μm apart. Experiments were conducted into the use of both photolithography and aerosol printing. For the latter, the mask was created using an Optomec AJ200 aerosol jet printer with a design created in AutoCAD 2018. An ink of PVDF-TrFE was prepared by dissolving PVDF-TrFE (70/30 weight%, Piezotech) in N-Methyl-2-pyrrolidone (NMP, Sigma-Aldrich) at 5 wt% under constant stirring at 60 °C. The stainless steel substrate was immobilised on the printer platen, with the platen temperature set at 40 °C. Approximately 20 mL of PVDF-TrFE ink was loaded into the bottle for the pneumatic atomizer. Using a 250 μm nozzle the ink was printed with: An atomizing flow rate of 1500 sscm, an exhaust flow rate of 1450 sscm and a sheath flow rate of 150 sscm. The design was followed 40 times in order to build up sufficient thickness to ensure mechanical stability. The substrate was then dried in an oven at 130 °C for 2 hours. Further details of this process are described in this previous work[34]. A 0.3 M $FeCl_3$ solution was used to etch regions where the resist/mask wasn't present (step 3). [35]$FeCl_3$ selectively switch to chrome as possible without degrading the structural integrity of the substrate) to enable gas (e.g. $H_2$) access to the anode. Solution etching rates are typically much higher than those of physical etching techniques such as reactive ion etching and ion milling, (the latter giving an etch rate of approximately 1 μm/hour for stainless steel [36]), making it the practical option for etching a 100 μm thick substrate. The disadvantage in this case is the isotropic nature of the etch. This can be mitigated by adopting a three-step procedure as demonstrated by Shimizu et al., though to date this has only been applied to much thinner substrates ~500nm) [37]. Samples were etched for approximately five hours at 70° C and stirred at 60 rpm for the first four hours.

X-ray diffraction measurements for samples on single-crystal substrates were conducted using a Panalytical Empyrean high resolution X-ray diffractometer using Cu-Kα radiation (λ = 1.5405 Å). Samples on the ABAD-buffered stainless steel were characterised using a Bruker D8 diffractometer (Cu-Kα radiation, λ = 1.5405 Å).

AFM measurements were performed on films using a Bruker Multimode 8 system in tapping mode. Commercial silicon cantilevers (Budget Sensors Ltd) with a resonance frequency of 300 kHz spring constant of 40 N/m were used to image 0.5 $μm^2$ areas at a scan frequency of 1 Hz. A conducting cobalt–chromium coated silicon cantilever with a resonance frequency of 150 kHz and a spring constant of 5 N/m was used for Kelvin probe force microscopy (KPFM) measurements.

Impedance measurements on the electrolyte structure were carried out using a probe station and hotplate together with a HP4294A impedance analyser. Regiments were conducted over a frequency range of 40 Hz – 1 MHz with a 50 mV AC voltage. The stainless steel substrate was grounded and a voltage applied to a circular Pt top electrode deposited by DC magnetron sputtering with a radius of approximately 250 μm defined by shadow mask.

Impedance spectroscopy measurements on the LSCF mesoporous cathode (deposited on single crystal YSZ) were carried out in a test station using Ag as a porous low impedance counter-electrode and a synthetic air atmosphere. Impedance spectra were acquired using an Alpha-A Novocontrol impedance spectrometer with ZG4 test interface, in a frequency range $10^6$-0.1 Hz and with a voltage amplitude of 0.05 V. For the fitting of the impedance data related to the cathode polarization arc, an equivalent circuit composed of a series of two ZARC elements (resistance in parallel to a CPE element) was used, being the simplest circuit able to fit the experimental data.

3. RESULTS AND DISCUSSION

The LSCF/MgO cathode was first characterised by XRD following the deposition on single-crystal (001) 8-YSZ substrates and the ABAD-buffered stainless steel substrates. In both cases, the films are highly oriented and epitaxial. **Fig. S1** shows X-ray diffractograms (XRD) of the LSCF/MgO film on (001) 8-YSZ. The film is oriented along the (110) axis as expected for growth of a perovskite material ($a_{LSCF}$ = 3.93Å) on 8-YSZ ($a_{YSZ}$ = 5.14Å)[38]. Here 3.93 x √2 = 5.54 giving a lattice mismatch of 7.2%, indicating domain matching epitaxy [13, 39]. A comparative XRD for the cathode grown on the ABAD-buffered stainless steel substrate (**Fig. S2**) (after **step 2** in Fig. 1) confirms the same (110) orientation of the LSCF, without second orientations. [40, 41][42][43, 44]XRD measurements of the etched films (after **Step 3**) (**Figs. S1 and S2** respectively for the single crystal and ABAD-buffered stainless steel substrates) confirm the MgO is removed by the etching process, leaving the LSCF structurally intact as discussed elsewhere [45].

AFM and TEM images showing film cross-sections before and after etching are shown in **Fig. 2.** **Fig. 2(a)** and **(b)** show etched columns over large and small areas from representative films grown on LSAT substrates. As with growth on 8-YSZ and ABAD-buffered stainless steel (**Figs. S3 and S4**), we observe a uniform mesoporous honeycomb matrix. The lateral feature dimensions are in the range ~30 nm (minimum distance observed ~15 nm, maximum distance ~40 nm), again similar to that on YSZ substrates (**Fig. S3**). **Fig. 2(c)** shows the composition of the columns from EDX analysis. The results demonstrate that the etching process is selective to the MgO, and that epitaxy of the LSCF is maintained, allowing a greater surface area for the oxygen reduction reaction to occur.

**Fig. 2(d)** and **(e)** show EDX maps before etching of a film on an LSAT substrate. The MgO regions are very clearly differentiated from the matrix. A high resolution TEM image of an etched film on an LSAT substrate (**Fig. 2 (f)**) shows a) the honeycomb retains the highly ordered perovskite matrix after etching, and b) the pore has a perfect bulb shape, the bulb forming owing to the poor wetting of the structural rocksalt MgO crystal to this and any of the substrates. The perovskite LCSF always forms the honeycomb matrix, and forms a seeding layer on the substrate, before the MgO region grows as columns with a bulbous region at the base owing to the high interfacial energy, and hence poor wetting, of the rock-salt MgO on the perovskite underlayer. The interfacial energy of the LSCF on the different substrates is always lower, even on YSZ (here the interfacial structural matching is provided by the (110) LCSF plane on the (100) perovskite [46-48][49][50]. **Fig. S6** shows further EDX analysis of the film of **Fig. 2(f)**, confirming all chemical elements present in the LSCF matrix, without Mg substitution, and only the Mg cation present in the MgO nanopillars.

**Fig. 2(g)** shows a KPFM image of a VAN film on ABAD-buffered stainless steel before removal of the MgO by etching. This confirms two distinct material phases in an approximately honeycomb structure of the matrix material. Further AFM images shown in **Fig. S3** show samples grown on single-crystal YSZ (with a $Ce_{0.8}Gd_{0.2}O_2$ (thickness ≈10 nm) buffer layer) before (**Fig. S3(a)**) and after (**Fig. S3(b)**) etching, while **Fig. S3** (c) and (d) show the equivalent images for films on ABAD-buffered stainless steel. These images confirm the removal of one material phase to leave a porous structure with feature sizes of similar size (after **step 3**), independent of the form of the YSZ substrate, whether ABAD or single crystal[51]. The equivalent 3D representations of these figures are given in the supplementary information (**Fig. S4). Fig. S5** (a) and (b) show additional top view SEM images of samples grown on single-crystal YSZ before and after etching respectively.

[13]Finally, **Fig. 2(h)** shows a TEM image of the etched film grown on ABAD-buffered stainless steel. The etched columns are ~15nm diameter, aligning with the columnar structure of the underlying YSZ and $CeO_2$. The selective removal of MgO upon etching is again confirmed by EDX analysis (not shown), from which we also observe some diffusion of Ce into the LSCF, as previously reported for LSCF electrodes grown on ceria-based electrolytes [52].

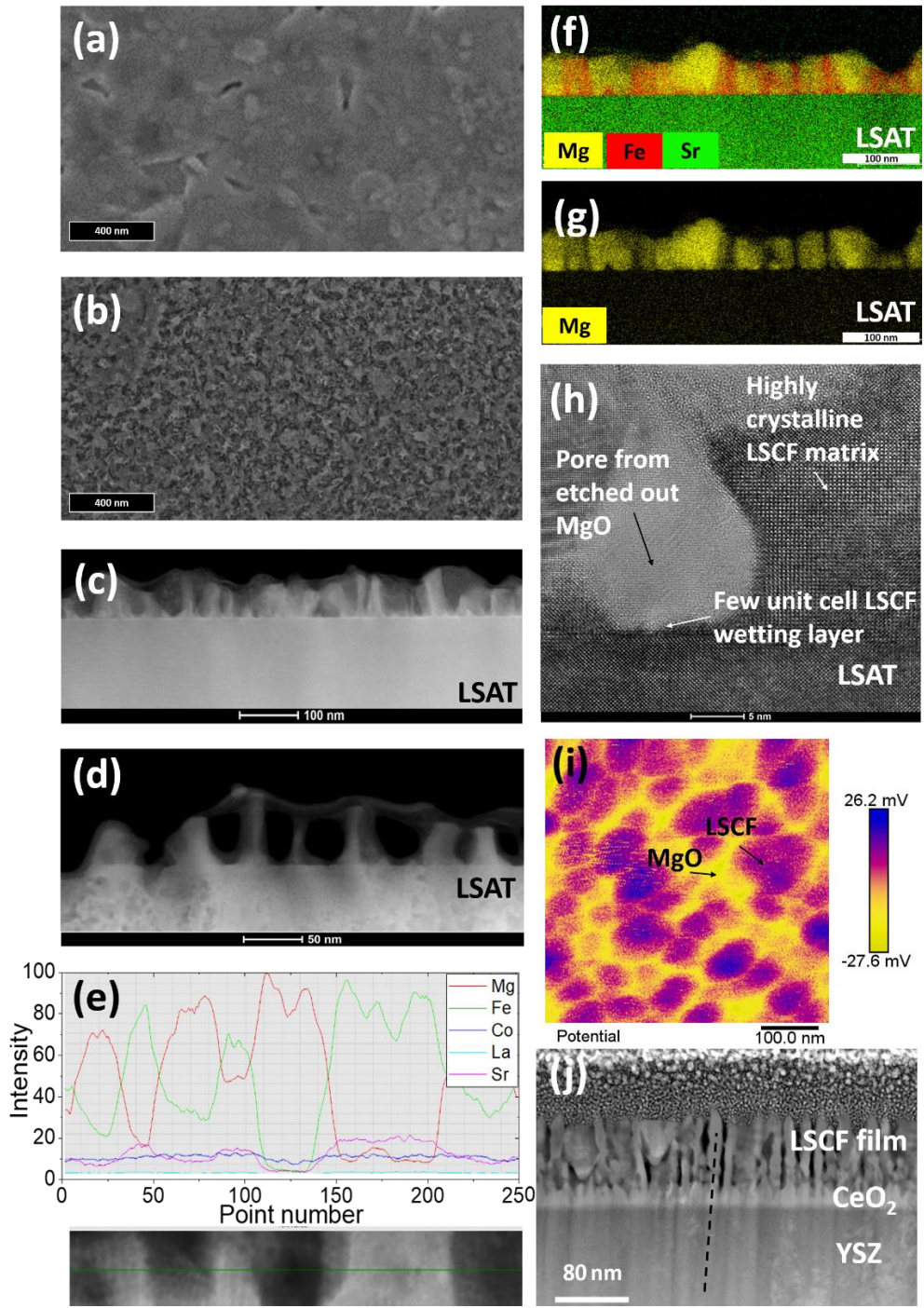

*Fig. 2: a), b) HAADF images of films grown on LSAT before and after etching c) d) EDX scan profile film before etching d), e) EDX measurements of LSCF/MgO film before etching of MgO f) TEM image of film after etching of MgO g) KPFM image of LSCF/MgO cathode grown on ABAD-buffered stainless steel h) TEM image of film grown on ABAD-buffered stainless steel after etching of MgO.*

We now explore the etching of the metallic substrate (after **Step 4** in **Fig. 1**). **Fig. 3(a)** shows a schematic of the mask pattern, with ~100 μm holes spaced ~500 μm apart, to be implemented by aerosol printing. **Fig. 3(b)** schematically shows complete etching with $FeCl_3$ giving cones of steel etched to the YSZ, and with the YSZ intact at the base of the cones. **Fig. 3(c)** shows an actual image viewed from the metal surface, after 5hr etching. Provisional experiments confirmed that the $FeCl_3$ terminated at the 15-YSZ interface, and that the electrolyte layer was able to self-support over a 300 x 100 µm$^2$ area. While fragile free-standing YSZ regions after the Ni etching did not retain perfection over the whole film area (**Fig. 3(c)**), the proof-of-concept of the simple etching process was demonstrated. Further improvements in the process are highly possible using several-step etching procedures [37]. Moreover, the film growth techniques demonstrated in this work could readily be implemented on novel substrates, for instance nanoporous Ni, on which virtually pore-free electrolytes can be deposited by PLD, thus allowing gas access without the need for etching and so further enhancing the commercial applicability of this work [2, 37].

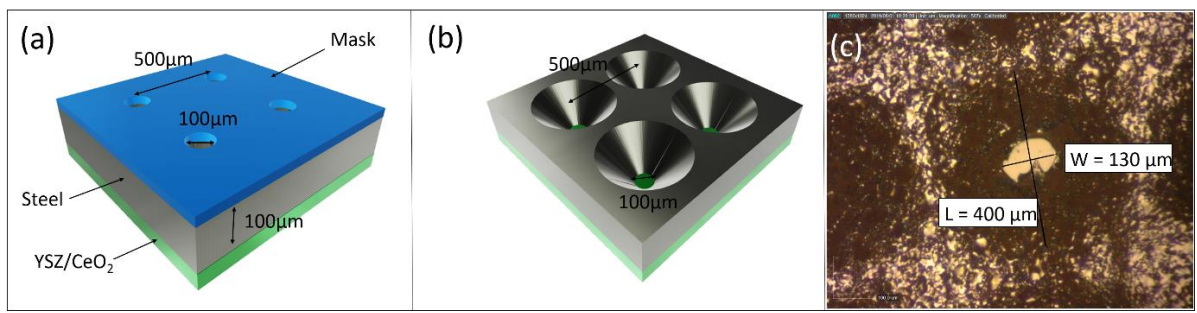

*Fig. 3: Schematics of etching process of **Step 4** showing (a) sample with mask applied before etching and (b) sample after etching. (c) 100µm holes etched using aerosol printed polymer mask - electrolyte layer collapsed*

Having demonstrated the growth of mesoporous VAN cathodes on both single-crystal and metal substrates and the ability to etch holes in the stainless steel substrate, we now turn to the electrical performance of the films. **Fig. 4(a)** shows the results of ionic conductivity measurements as a function of temperature for the $CeO_2$/YSZ films on stainless steel determined from electrochemical impedance spectroscopy (EIS) (**Fig. S7**). The data were acquired between 275° C and 480° C to approximate the low-temperature operating regime of a µSOFC. The ionic conductivity and $E_a$ (0.79 eV) are comparable to literature values, thus showing the high quality of the films, and confirming the validity of this approach for the fabrication of µSOFCs [53].

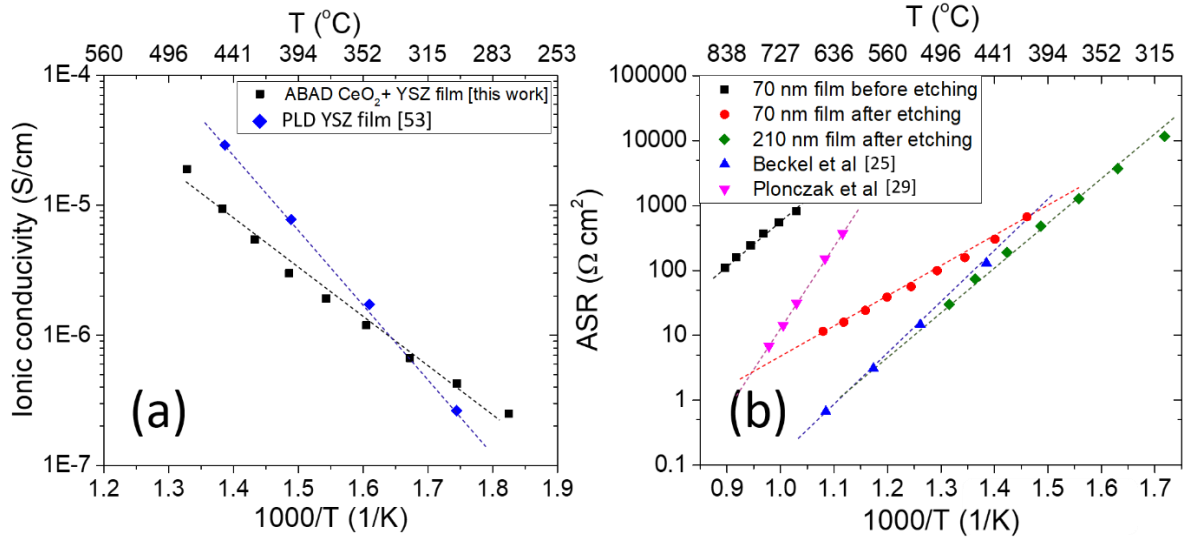

*Fig. 4: Electrical data of electrolyte and cathode layers. a) Arrhenius plot of ionic conductivity of YSZ films deposited by ABAD on stainless steel [53]; b) Area specific resistance of LSCF/MgO film before and after etching of MgO as compared to literature results from Beckel et al. and Plonczak et al. [25, 29]*

The area specific resistance (ASR) of the thin film cathodes was measured by growing the thin-film on one side of a YSZ substrate (with a $Ce_{0.8}Gd_{0.2}O_2$ (thickness ≈10 nm) buffer layer to reduce substrate – film lattice mismatch and to avoid the formation of secondary phases at the interface) and depositing a low impedance, porous Ag counter electrode on the other (see Supplementary **Figure S**8 for representative Nyquist plots and impedance fittings). In this way, the impedance response is dominated by the thin-film cathode [54]. **Fig. 4(b)** shows the ASR of a 70 nm film measured as a function of temperature before and after the MgO was removed by etching in acetic acid. Before etching, the presence of the MgO (which is inert towards ORR [56]) results in ASR values of approximately 100 $\Omega cm^2$ at 840° C, several orders of magnitude higher than are observed for plain LSCF films in the literature (as shown by the pink and blue lines in **Fig. 4(b)**). Additionally, **Fig. 4(b)** shows the ASR values of etched films with thickness 70 nm and 210 nm. The increased ASR in the unetched films can be ascribed to two factors associated with the VAN films:

*a)* the MgO incorporated in the film is highly resistive and reduces the areal fraction of the LSCF;

*b)* as determined below, the as-grown VAN film is under *out-of-plane* tension: such tensile strain is known to induce Sr surface segregation [38, 57], resulting in electronically and ionically insulating Sr-enriched phases, which severely impedes surface oxygen exchange [58]. [59] An *out-of-plane* tension in the as-grown VAN film is expected because of the presence of stiff, and larger lattice parameter MgO *cf*. LSCF (i.e. 4.21 Å compared to ~3.8-3.9 Å [38, 49]). The strain state in the VAN film before and after etching was determined by XRD. For a VAN film on a YSZ single crystal substrate, **Fig. S1** shows a shift in X-ray peak position of the LSCF (110) peak to give (a+b)/2 = 3.922 ± 0.002 Å before etching to 3.913 ± 0.002 Å after etching. This indicates a tensile strain state which is relaxed after etching out the MgO, as shown in the schematic of **Fig. 5**. The same level of strain relaxation was also observed on the ABAD-buffered stainless steel. Since the lattice parameters of LSCF are dependent upon its precise composition varying by around 0.1Å for different cation stoichiometries, it is not possible to ascertain the level of strain compared to bulk LSCF [38].

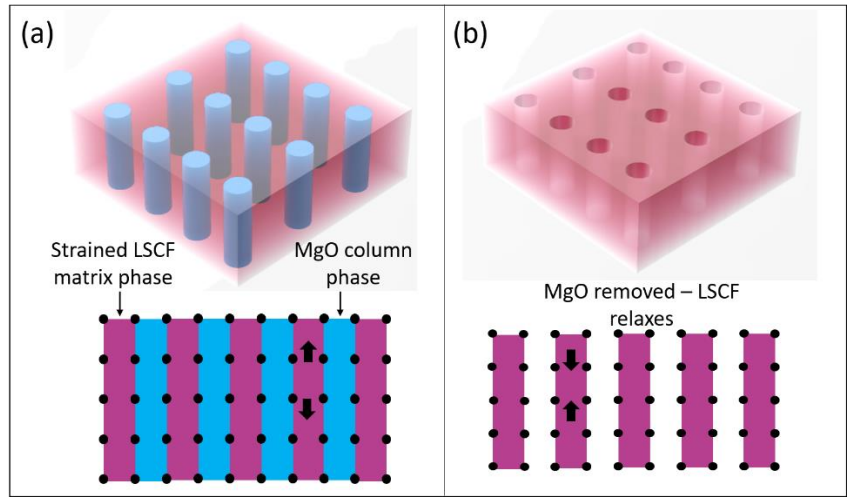

*Fig. 5: (a) LSCF/MgO VAN film, and (b) strain relaxation after etching MgO*

After etching the 70 nm film, a substantial decrease is observed in both the value of ASR, now approximately 100 $\Omega cm^2$ at 500° C, and the activation energy ($E_a$) (from 1.6 ± 0.1 eV before etching to 0.97 ± 0.01 eV after etching), leading, at < ~400° C, to lower values of ASR than for plain LSCF films.

[60]While a decrease in ASR upon etching the 70 nm could be in principle explained by a higher surface area available for solid-gas reaction, the lower $E_a$ value suggests a more subtle change in the oxygen incorporation kinetics upon etching. Particularly, this could be rationalized by the appearance of a surface oxide ion transport path, parallel to the bulk path, in the porous structure[61]. Besides, one can conside[61]r chemical compositional effects associated with the tensile strain and etching process. As already noted, tensile strain leads to Sr segregation which would be reduced after etching. A change in surface stoichiometry of the LSCF by the etching process is also a possibility. In fact, as observed from Fig. 2h, the LSCF matrix is highly intact with no reaction products visible on the LSCF surface after etching. The atomic scale information about the chemical etching effect is beyond the scope of this study and is dealt with in other separate studies[45].

Comparing the ASR values for etched 70 nm and 210 nm films, the 210 nm film offers ~1 $\Omega cm^2$ at 600° C and $E_a$~1.5 eV, in line with state-of-the-art performance of porous LSCF[25]. For temperatures <400° C however, the lower activation energy of the 70 nm film once again yields lower ASR. This result is important for μSOFCs for operation below 400° C. The different activation energy for the 70 nm etched film with respect to the 210 nm etched film again suggests the presence of two different oxide ion transport paths (bulk and surface), the latter becoming predominant for low electrode thicknesses[61]. A further detailed electrochemical study will be conducted next to further elucidate the effect of film thickness on the dominant transport path.

## 4. CONCLUSIONS

We have shown that it is possible to grow state-of-the-art epitaxial, mesoporous LSCF thin film cathodes on commercially available stainless steel metallic substrates with biaxially aligned YSZ electrolyte layers. Firstly, we demonstrated that the commercially available YSZ electrolytes on stainless steel feature $3 \times 10^{-6}$ S.cm$^{-1}$ ionic conductivity at 400° C. Such performance is consistent with typical high-performance materials used for electrochemical devices. Thereafter, we showed that single crystalline epitaxial and porous LSCF cathodes feature area specific resistance of 100 Ωcm at 500° C and an activation energy of 0.97 eV, demonstrating state-of-the-art performance. Finally, we showed that it is possible to pattern and chemically etch holes of ~ 100 μm dimension into the μSOFC device for allowing gas access. Our work marks an important advance towards creating high-performance μSOFCs for portable power applications via a proven commercial metallic substrate system.

ASSOCIATED CONTENT

The following files are available free of charge

XRD measurements of LSCF/MgO cathode on single-crystal YSZ (TIF)

XRD measurements of YSZ (1500 nm) and CeO$_2$ (50 nm) buffered stainless steel substrates (TIF)

AFM images of LSCF/MgO films on single-crystal YSZ before and after etching of MgO, and of LSCF/MgO films on ABAD-buffered stainless steel before and after etching of MgO (TIF)

3D representation of AFM images of LSCF/MgO films on single-crystal YSZ before and after etching of MgO, and of LSCF/MgO films on ABAD YSZ-buffered stainless steel before and after etching of MgO (TIF)

SEM images of LSCF/MgO film on single-crystal 8-YSZ before and after etching (TIF)

HAADF and EDX images of etched LSCF film on LSAT, confirming phase separation and film composition (TIF)

EIS measurements of YSZ (1500 nm) and CeO$_2$ (50 nm) buffered stainless steel substrates (TIF)

Representative Nyquist plots for LSCF-MgO before and after etching (TIF)

AUTHOR INFORMATION

**Corresponding Author**

*mpw52@cam.ac.uk11

## Author Contributions

The manuscript was written through contributions of all authors. All authors have given approval to the final version of the manuscript. Samples were grown by MW and MA; XRD and AFM measurements conducted by MW and MA; EIS measurements done by MW and AL; ASR measurements done by FB; TEM and EDX performed by HW, JD, and XW; aerosol printing conducted by TC.

## Funding Sources

We acknowledge substrate provision from Bruker HTS, via our H2020 project Eurotapes, EUROTAPES, a collaborative project funded by the European Commission's Seventh Framework Program (FP7/2007–2013) under Grant Agreement No. 280432.

JLM-D and MPW acknowledge the ERC POC grant, Portapower,779444. JLM-D also acknowledges support from the Royal Academy of Engineering Chair in Emerging technologies grant CIET1819_24, and the EPSRC Centre of Advanced Materials for Integrated Energy Systems (CAM-IES), grant EP/P007767/1.

M.A. acknowledges the support from the Feodor Lynen Research Fellowship Program of the Alexander von Humboldt Foundation and the Isaac Newton Trust, Minute

M.A. acknowledges the support from the Feodor Lynen Research Fellowship Program of the Alexander von Humboldt Foundation the Isaac Newton Trust 17.25(a) and support from the Centre of Advanced Materials for Integrated Energy Systems (CAM-IES) under EP/P007767/1.

SK-N acknowledges support from the European Research Council through an ERC Starting Grant (ERC-2014-STG-639526, NANOGEN).

The TEM/HRTEM work at Purdue University was supported by the U.S. National Science Foundation (DMR-1565822 and DMR-2016453).

Part of this project has received funding from the European Union's Horizon 2020 research and innovation program under grant agreement No 824072 (HARVESTORE) and No 681146 (ULTRASOFC) and was supported by an STSM Grant from the COST Action MP1308: Towards Oxide-Based Electronics (TO-BE), supported by COST (European Cooperation in Science and Technology).

REFERENCES

1. Bieberle-Hutter, A., et al., *A micro-solid oxide fuel cell system as battery replacement.* Journal of Power Sources, 2008. **177**(1): p. 123-130.
2. Lee, Y., Y.M. Park, and G.M. Choi, *Micro-solid oxide fuel cell supported on a porous metallic Ni/stainless-steel bi-layer.* Journal of Power Sources, 2014. **249**: p. 79-83.
3. Kwon, C.W., et al., *High-Performance Micro-Solid Oxide Fuel Cells Fabricated on Nanoporous Anodic Aluminum Oxide Templates.* Advanced Functional Materials, 2011. **21**(6): p. 1154-1159.
4. Pihlatie, M., T. Ramos, and A. Kaiser, *Testing and improving the redox stability of Ni-based solid oxide fuel cells.* Journal of Power Sources, 2009. **193**(1): p. 322-330.
5. Al-Khori, K., Y. Bicer, and M. Koc, *Integration of Solid Oxide Fuel Cells into oil and gas operations: needs, opportunities, and challenges.* Journal of Cleaner Production, 2020. **245**: p. 18.
6. Garbayo, I., et al., *Full ceramic micro solid oxide fuel cells: towards more reliable MEMS power generators operating at high temperatures.* Energy & Environmental Science, 2014. **7**(11): p. 3617-3629.
7. Huang, H., et al., *High-performance ultrathin solid oxide fuel cells for low-temperature operation.* Journal of the Electrochemical Society, 2007. **154**(1): p. B20-B24.
8. Watanabe, H., N. Yamada, and M. Okaji, *Linear thermal expansion coefficient of silicon from 293 to 1000 K.* International Journal of Thermophysics, 2004. **25**(1): p. 221-236.
9. Yamamoto, O., *Solid oxide fuel cells: fundamental aspects and prospects.* Electrochimica Acta, 2000. **45**(15-16): p. 2423-2435.
10. Baertsch, C.D., et al., *Fabrication and structural characterization of self-supporting electrolyte membranes for a micro solid-oxide fuel cell.* Journal of Materials Research, 2004. **19**(9): p. 2604-2615.
11. Acosta, M., et al., *Nanostructured Materials and Interfaces for Advanced Ionic Electronic Conducting Oxides.* Advanced Materials Interfaces, 2019. **6**(15).
12. Santiso, J. and M. Burriel, *Deposition and characterisation of epitaxial oxide thin films for SOFCs.* Journal of Solid State Electrochemistry, 2011. **15**(5): p. 985-1006.
13. Mori, D., et al., *Synthesis, structure, and electrochemical properties of epitaxial perovskite La0.8Sr0.2CoO3 film on YSZ substrate.* Solid State Ionics, 2006. **177**(5-6): p. 535-540.
14. Sanna, S., et al., *Fabrication and Electrochemical Properties of Epitaxial Samarium-Doped Ceria Films on SrTiO3-Buffered MgO Substrates.* Advanced Functional Materials, 2009. **19**(11): p. 1713-1719.
15. Shao, Z.P. and S.M. Haile, *A high-performance cathode for the next generation of solid-oxide fuel cells.* Nature, 2004. **431**(7005): p. 170-173.
16. Zomorrodian, A., et al., *Electrical conductivity of epitaxial La0.6Sr0.4Co0.2Fe0.8O3-delta thin films grown by pulsed laser deposition.* International Journal of Hydrogen Energy, 2010. **35**(22): p. 12443-12448.
17. Joo, J.H. and G.M. Choi, *Simple fabrication of micro-solid oxide fuel cell supported on metal substrate.* Journal of Power Sources, 2008. **182**(2): p. 589-593.
18. Vafaeenezhad, S., et al., *Development of proton conducting fuel cells using nickel metal support.* Journal of Power Sources, 2019. **435**.

19. Rey-Mermet, S., et al., *Nanoporous YSZ film in electrolyte membrane of Micro-Solid Oxide Fuel Cell.* Thin Solid Films, 2010. **518**(16): p. 4743-4746.
20. Jiang, J., et al., *Improved ionic conductivity in strained yttria-stabilized zirconia thin films.* Applied Physics Letters, 2013. **102**(14): p. 4.
21. Yang, S.M., et al., *Strongly enhanced oxygen ion transport through samarium-doped $CeO_2$ nanopillars in nanocomposite films.* Nature Communications, 2015. **6**.
22. Lee, S. and J.L. MacManus-Driscoll, *Research Update: Fast and tunable nanoionics in vertically aligned nanostructured films.* Apl Materials, 2017. **5**(4).
23. Lee, S., et al., *Ionic Conductivity Increased by Two Orders of Magnitude in Micrometer-Thick Vertical Yttria-Stabilized $ZrO_2$ Nanocomposite Films.* Nano Letters, 2015. **15**(11): p. 7362-7369.
24. Garbayo, I., et al., *Thin film oxide-ion conducting electrolyte for near room temperature applications.* Journal of Materials Chemistry A, 2019. **7**(45): p. 25772-25778.
25. Beckel, D., et al., *Electrochemical performance of LSCF based thin film cathodes prepared by spray pyrolysis.* Solid State Ionics, 2007. **178**(5-6): p. 407-415.
26. Tsvetkov, N., et al., *Improved chemical and electrochemical stability of perovskite oxides with less reducible cations at the surface.* Nature Materials, 2016. **15**(9): p. 1010-1016.
27. Riva, M., et al., *Influence of surface atomic structure demonstrated on oxygen incorporation mechanism at a model perovskite oxide.* Nature Communications, 2018. **9**.
28. Yoon, J., et al., *Vertically Aligned Nanocomposite Thin Films as a Cathode/Electrolyte Interface Layer for Thin-Film Solid-Oxide Fuel Cells.* Advanced Functional Materials, 2009. **19**(24): p. 3868-3873.
29. Plonczak, P., et al., *Electrochemical Characterization of $La_{0.58}Sr_{0.4}Co_{0.2}Fe_{0.8}O_3$-delta Thin Film Electrodes Prepared by Pulsed Laser Deposition.* Journal of the Electrochemical Society, 2012. **159**(5): p. B471-B482.
30. Obradors, X. and T. Puig, *Coated conductors for power applications: materials challenges.* Superconductor Science & Technology, 2014. **27**(4): p. 17.
31. Usoskin, A., et al., *Long HTS Coated Conductor Processed via Large-Area PLD/ABAD for High-Field Applications.* Ieee Transactions on Applied Superconductivity, 2016. **26**(3): p. 4.
32. Buckeridge, J., et al., *Dynamical response and instability in ceria under lattice expansion.* Physical Review B, 2013. **87**(21): p. 10.
33. Niu, G., et al., *On the local electronic and atomic structure of $Ce_{1-x}Pr_xO_2$-delta epitaxial films on Si.* Journal of Applied Physics, 2014. **116**(12): p. 9.
34. Smith, M., et al., *Controlling and assessing the quality of aerosol jet printed features for large area and flexible electronics.* Flexible and Printed Electronics, 2017. **2**(1).
35. Chen, X., et al., *Thin-film heterostructure solid oxide fuel cells.* Applied Physics Letters, 2004. **84**(14): p. 2700-2702.
36. Nishi, Y., et al., *ION MILLING OF NITRIZED 18-8 STAINLESS-STEEL.* Journal of Materials Science Letters, 1987. **6**(1): p. 63-64.
37. Shimizu, M., et al., *Anisotropic multi-step etching for large-area fabrication of surface microstructures on stainless steel to control thermal radiation.* Science and Technology of Advanced Materials, 2015. **16**(2).
38. Yu, Y., et al., *Effect of Sr Content and Strain on Sr Surface Segregation of $La_{1-x}Sr_xCo_{0.2}Fe_{0.8}O_3$-delta as Cathode Material for Solid Oxide Fuel Cells.* Acs Applied Materials & Interfaces, 2016. **8**(40): p. 26704-26711.

39. Smith, C.R., et al., *Structural Investigation of Perovskite Manganite and Ferrite Films on Yttria-Stabilized Zirconia Substrates.* Journal of the Electrochemical Society, 2012. **159**(8): p. F436-F441.
40. Choi, Y., M.e.C. Lin, and M. Liu, *Computational study on the catalytic mechanism of oxygen reduction on La0. 5Sr0. 5MnO3 in solid oxide fuel cells.* Angewandte Chemie International Edition, 2007. **46**(38): p. 7214-7219.
41. Wang, Z., et al., *Oxygen reduction and transport on the La 1− x Sr x Co 1− y Fe y O 3− δ cathode in solid oxide fuel cells: a first-principles study.* Journal of Materials Chemistry A, 2013. **1**(41): p. 12932-12940.
42. Develos-Bagarinao, K., et al., *Oxygen surface exchange properties and surface segregation behavior of nanostructured La 0.6 Sr 0.4 Co 0.2 Fe 0.8 O 3− δ thin film cathodes.* Physical Chemistry Chemical Physics, 2019. **21**(13): p. 7183-7195.
43. Chen, A., et al., *Microstructure, vertical strain control and tunable functionalities in self-assembled, vertically aligned nanocomposite thin films.* Acta Materialia, 2013. **61**(8): p. 2783-2792.
44. Huang, J., J.L. MacManus-Driscoll, and H. Wang, *New epitaxy paradigm in epitaxial self-assembled oxide vertically aligned nanocomposite thin films.* Journal of Materials Research, 2017. **32**(21): p. 4054-4066.
45. Acosta, M., et al., *Ultrafast oxygen reduction kinetics in (La,Sr)(Co,Fe)O3 vertically aligned nanocomposites below 400 °C.* Submitted 2020, Nature Materials.
46. MacManus-Driscoll, J.L., *Self-assembled heteroepitaxial oxide nanocomposite thin film structures: designing interface-induced functionality in electronic materials.* Advanced Functional Materials, 2010. **20**(13): p. 2035-2045.
47. Sun, X., J.L. MacManus-Driscoll, and H. Wang, *Spontaneous ordering of oxide-oxide epitaxial vertically aligned nanocomposite thin films.* Annual Review of Materials Research, 2019. **50**.
48. MacManus-Driscoll, J.L., et al., *Strain control and spontaneous phase ordering in vertical nanocomposite heteroepitaxial thin films.* Nature materials, 2008. **7**(4): p. 314-320.
49. *Magnesium oxide (MgO) crystal structure, lattice parameters, thermal expansion: Datasheet from Landolt-Börnstein - Group III Condensed Matter · Volume 41B: "II-VI and I-VII Compounds; Semimagnetic Compounds" in SpringerMaterials (https://doi.org/10.1007/10681719_206)*, O. Madelung, U. Rössler, and M. Schulz, Editors., Springer-Verlag Berlin Heidelberg.
50. Jayaraj, M., *Nanostructured Metal Oxides and Devices.* Springer.
51. Singh, S., et al., *Growth of Doped SrTiO3 Ferroelectric Nanoporous Thin Films and Tuning of Photoelectrochemical Properties with Switchable Ferroelectric Polarization.* ACS Applied Materials & Interfaces, 2019. **11**(49): p. 45683-45691.
52. Giannici, F., et al., *Cation Diffusion and Segregation at the Interface between Samarium-Doped Ceria and LSCF or LSFCu Cathodes Investigated with X-ray Microspectroscopy.* Acs Applied Materials & Interfaces, 2017. **9**(51): p. 44466-44477.
53. Jiang, J. and J.L. Hertz, *On the variability of reported ionic conductivity in nanoscale YSZ thin films.* Journal of Electroceramics, 2014. **32**(1): p. 37-46.
54. Rupp, G.M., et al., *Real-time impedance monitoring of oxygen reduction during surface modification of thin film cathodes.* Nature Materials, 2017. **16**(6): p. 640-+.

55. Shen, F. and K. Lu, *Comparison of Different Perovskite Cathodes in Solid Oxide Fuel Cells.* Fuel Cells, 2018. **18**(4): p. 457-465.
56. Sempolinski, D.R. and W.D. Kingery, *IONIC-CONDUCTIVITY AND MAGNESIUM VACANCY MOBILITY IN MAGNESIUM-OXIDE.* Journal of the American Ceramic Society, 1980. **63**(11-1): p. 664-669.
57. Chevalier, S., et al., *Effect of nano-layered ceramic coatings on the electrical conductivity of oxide scale grown on ferritic steels.* Journal of Applied Electrochemistry, 2009. **39**(4): p. 529-534.
58. Hess, F. and B. Yildiz, *Polar or not polar? The interplay between reconstruction, Sr enrichment, and reduction at the La0.75Sr0.25MnO3 (001) surface.* Physical Review Materials, 2020. **4**(1).
59. Jung, W. and H.L. Tuller, *Investigation of surface Sr segregation in model thin film solid oxide fuel cell perovskite electrodes.* Energy & Environmental Science, 2012. **5**(1): p. 5370-5378.
60. Im, H.-N., et al., *Investigation of oxygen reduction reaction on La0. 1Sr0. 9Co0. 8Fe0. 2O3-δ electrode by electrochemical impedance spectroscopy.* Journal of the Electrochemical Society, 2015. **162**(7): p. F728.
61. Lu, Y.X., C. Kreller, and S.B. Adler, *Measurement and Modeling of the Impedance Characteristics of Porous La1-xSrxCoO3-delta Electrodes.* Journal of the Electrochemical Society, 2009. **156**(4): p. B513-B525.